\documentstyle [aps,psfig]{revtex}
\begin{document}
\draft
\title{{\bf Measurement of the form-factor ratios for 
$D^+_s\rightarrow \phi \,\ell^+ \nu_\ell$}}
%
%

%
\author{
    E.~M.~Aitala,$^9$
       S.~Amato,$^1$
    J.~C.~Anjos,$^1$
    J.~A.~Appel,$^5$
       D.~Ashery,$^{14}$
       S.~Banerjee,$^5$
       I.~Bediaga,$^1$
       G.~Blaylock,$^8$
    S.~B.~Bracker,$^{15}$
    P.~R.~Burchat,$^{13}$
    R.~A.~Burnstein,$^6$
       T.~Carter,$^5$
 H.~S.~Carvalho,$^{1}$
  N.~K.~Copty,$^{12}$
    L.~M.~Cremaldi,$^9$
 C.~Darling,$^{18}$
       K.~Denisenko,$^5$
       A.~Fernandez,$^{11}$
       G.F.~Fox,$^{12}$
       P.~Gagnon,$^2$
       C.~Gobel,$^1$
       K.~Gounder,$^9$
     A.~M.~Halling,$^5$
       G.~Herrera,$^4$
 G.~Hurvits,$^{14}$
       C.~James,$^5$
    P.~A.~Kasper,$^6$
       S.~Kwan,$^5$
    D.~C.~Langs,$^{12}$
       J.~Leslie,$^2$
       B.~Lundberg,$^5$
       S.~MayTal-Beck,$^{14}$
       B.~Meadows,$^3$
       J.~R.~T.~de~Mello~Neto,$^1$
       D.~Mihalcea,$^7$
    R.~H.~Milburn,$^{16}$
 J.~M.~de~Miranda,$^1$
       A.~Napier,$^{16}$
       A.~Nguyen,$^7$
  A.~B.~d'Oliveira,$^{3,11}$
       K.~O'Shaughnessy,$^2$
    K.~C.~Peng,$^6$
    L.~P.~Perera,$^3$
    M.~V.~Purohit,$^{12}$
       B.~Quinn,$^9$
       S.~Radeztsky,$^{17}$
       A.~Rafatian,$^9$
    N.~W.~Reay,$^7$
    J.~J.~Reidy,$^9$
    A.~C.~dos Reis,$^1$
    H.~A.~Rubin,$^6$
    D.~A.~Sanders,$^9$
 A.~K.~S.~Santha,$^3$
 A.~F.~S.~Santoro,$^1$
       A.~J.~Schwartz,$^3$
       M.~Sheaff,$^{4,17}$
    R.~A.~Sidwell,$^7$
    A.~J.~Slaughter,$^{18}$
    M.~D.~Sokoloff,$^3$
    J.~Solano,$^1$
       N.~R.~Stanton,$^7$
    R.~J.~Stefanski,$^5$   
       K.~Stenson,$^{17}$
    D.~J.~Summers,$^9$
 S.~Takach,$^{18}$
       K.~Thorne,$^5$
    A.~K.~Tripathi,$^{7}$
       S.~Watanabe,$^{17}$
 R.~Weiss-Babai,$^{14}$
       J.~Wiener,$^{10}$
       N.~Witchey,$^7$
       E.~Wolin,$^{18}$
       S.M.~Yang,$^7$
       D.~Yi,$^9$
       S.~Yoshida,$^{7}$                         
       R.~Zaliznyak,$^{13}$
       and
       C.~Zhang$^7$ \\
(Fermilab E791 Collaboration)
}

\address{
$^1$ Centro Brasileiro de Pesquisas F\'\i sicas, Rio de Janeiro, Brazil\\
$^2$ University of California, Santa Cruz, California 95064\\
$^3$ University of Cincinnati, Cincinnati, Ohio 45221\\
$^4$ CINVESTAV, Mexico\\
$^5$ Fermilab, Batavia, Illinois 60510\\
$^6$ Illinois Institute of Technology, Chicago, Illinois 60616\\
$^7$ Kansas State University, Manhattan, Kansas 66506\\
$^8$ University of Massachusetts, Amherst, Massachusetts 01003\\
$^9$ University of Mississippi, University, Mississippi 38677\\
$^{10}$ Princeton University, Princeton, New Jersey 08544\\
$^{11}$ Universidad Autonoma de Puebla, Mexico\\
$^{12}$ University of South Carolina, Columbia, South Carolina 29208\\
$^{13}$ Stanford University, Stanford, California 94305\\
$^{14}$ Tel Aviv University, Tel Aviv, Israel\\
$^{15}$ Box 1290, Enderby, BC, VOE 1V0, Canada\\
$^{16}$ Tufts University, Medford, Massachusetts 02155\\
$^{17}$ University of Wisconsin, Madison, Wisconsin 53706\\
$^{18}$ Yale University, New Haven, Connecticut 06511\\
}

\maketitle
\today

\begin{abstract}
We have measured
the form factor ratios $r_V = V(0)/A_1(0)$ and $r_2 = A_2(0)/A_1(0)$
for the decay $D_s^+ \rightarrow \phi \, \ell^+ \, \nu_\ell$,
$\phi \rightarrow K^+ \, K^-$, using data from
charm hadroproduction experiment E791 at Fermilab. Results are
based on 144 signal and 22 background events in the electron channel
and 127 signal and 34 background events in the muon channel.
We combine the measurements from both lepton channels to obtain 
$r_V = 2.27 \pm 0.35 \pm 0.22$ and $r_2 = 1.57 \pm 0.25 \pm 0.19$.

\end{abstract}
\narrowtext

Semileptonic decays of mesons containing one heavy quark are described in
simplest terms by a spectator model in which the heavy quark decays
semileptonically while the light spectator antiquark passes from the parent
meson to the final-state meson.  These decays are still relatively simple
when one considers hadrons rather than quarks;
the leptonic and hadronic currents factorize, and all strong-interaction
effects can be parametrized by a few form factors~\cite{KS}.

If SU(3) flavor symmetry is approximately valid, replacing a
spectator $\overline{d}$ quark by a spectator  $\overline{s}$ quark should have
little effect on the form factors. We have investigated this
spectator replacement experimentally in charm meson decay, and report here new
measurements by Fermilab experiment E791 of form factor ratios  for $D_s^+
\rightarrow \phi\, e^+\, \nu_e$ and $D_s^+ \rightarrow \phi\, \mu^+\, \nu_\mu$,
with $\phi \rightarrow K^+\, K^-$.
These $D_s$ results are then combined and
compared  to recent high-statistics form-factor ratio
results~\cite{renata,daniel}
for $D^+ \rightarrow \overline{K}^{\,*0}\, \ell^+ \,\nu_\ell$ measured in the
same experiment, and also to theoretical predictions. (Charge-conjugate states
are also included by inference throughout this letter.)

In addition to testing SU(3) symmetry in semileptonic decays, ratios of form
factors in $D_s^+ \rightarrow \phi\,\ell^+\, \nu_\ell$ are needed for tests
of factorization in $D_s$ hadronic decays~\cite{KAMAL}.

With a vector meson in the final state, there are four form factors, $V(q^2)$,
$A_1(q^2)$, $A_2(q^2)$ and $A_3(q^2)$, which are functions of the
Lorentz-invariant momentum transfer squared~\cite{KS}. The limited size of
current data samples precludes precise measurement of the $q^2$-dependence of
form factors, and thus we assume a nearest-pole dominance model: $F(q^2) =
F(0)/(1-q^2/m_{pole}^2)$ where $m_{pole} = m_V = 2.1 \: {\rm GeV}/c^2$ for
the vector form factor $V$, and $m_{pole} = m_A = 2.5 \: {\rm GeV}/c^2$ for
the axial-vector form factors~\cite{PDG94}. Because $A_1(q^2)$ appears among
the coefficients of every term in the differential decay rate, it is
customary to factor out $A_1(0)$  and to measure the ratios $r_V =
V(0)/A_1(0)$, $r_2 = A_2(0)/A_1(0)$ and $r_3 = A_3(0)/A_1(0)$ which are
independent of the absolute normalization of the data.

The form factor $A_3$, which probes the spin-0 component of the virtual
$W$, is unobservable in the limit of vanishing lepton mass, and has been
found to be small in $D^+ \rightarrow \overline{K}^{\,*0} \mu^+ \,
\nu_\mu$~\cite{daniel}. However, the $q^2$ dependence of the muon channel
is more complex than that of the electron channel even if $A_3$ vanishes.
The decay rate at low $q^2$ is suppressed by an overall factor
$(1-m_\mu^2/q^2)^2$, and additional spin-flip amplitudes, suppressed by an
overall factor of $m_{\mu}^2/q^2$ relative to the spin non-flip amplitudes,
contribute to the differential decay rate at small $q^2$. E791 is the first
experiment to measure form factor ratios for $D_s^+$ semileptonic
decays in both electron and muon modes. Obtaining consistent results
for the two lepton channels gives confidence that these $q^2$
differences are adequately understood.

Fermilab E791 is a fixed-target charm hadroproduction experiment~\cite{APPEL}.
Charm particles were produced in collisions of a $500 \: {\rm GeV}/c$
$\pi^-$ beam with five thin targets, one platinum and four diamond. About $2
\times 10^{10}$ events were recorded during the 1991-1992 Fermilab
fixed-target run. The tracking system consisted of 23 planes of silicon
microstrip detectors, 45 planes of drift and proportional wire chambers, and
two large-aperture dipole magnets. Hadron identification is based on the
information from two multicell \v Cerenkov counters that provide good
discrimination between kaons and pions in the momentum range $6 - 36 \: {\rm
GeV}/c$. In this momentum range, the probability of misidentifying a pion as
a kaon depends on momentum but does not exceed 5\%. 

Electron candidates are identified by electromagnetic shower shape, the match
between calorimeter energy and tracking momentum, and agreement between
calorimeter and tracking position measurements. Muon candidates are
identified by a single plane of scintillator strips, oriented horizontally,
located behind an equivalent of 2.4 meters of iron (comprised of the
calorimeters and one meter of bulk steel shielding). The angular acceptance of
the scintillator plane was about $\pm 62 \:{\rm mrad} \times \pm 48 \:{\rm
mrad}$ (horizontally and vertically, respectively), which was somewhat smaller
than that of the rest of the spectrometer for tracks that went through both
magnets ($\approx\!\pm 100 \:{\rm mrad} \times \pm 64 \:{\rm mrad}$). The
vertical position of a scintillator hit is determined from the strip's
vertical position, and the horizontal position of a hit from timing
information.

The event selection criteria used for this analysis are the same for both
electron and muon channels except for those related to lepton identification.
Events are selected if they contain an acceptable decay vertex determined by
the intersection point of three tracks that have been identified as a lepton
and two kaons. The longitudinal separation between this candidate decay vertex
and the reconstructed production vertex is required to be at least 11 times the
estimated error on the separation, and the candidate decay vertex must occur
outside the target material.

We require electron candidates to have a minimum momentum of 5~GeV/$c$,
which gives an electron identification probability of about 70\% and
a probabability for a pion to be misidentified as an electron of 1-2\%.
We retain only muon candidates with momenta greater than 8 GeV/c
to reduce the contamination from hadron decays in flight.
With this
momentum restriction, the muon identification efficiency is about 85\%, and
the probability for a hadron to be misidentified as a muon is about 3\%.

Events in which the lepton candidate has the same charge as the candidate
decay are assigned to a ``right-sign" (RS) sample. This sample contains both
signal events and a small amount of background  arising from reconstruction
errors, including misreconstructed charm, and possible kinematic reflections
from other charm decay channels. Events in which the lepton candidate and the
candidate decay have opposite charge are assigned to a ``wrong-sign" (WS)
channel, which contains purely background. Because WS events have two hadrons
of the same charge, this sample is dominated by reconstruction background. We
expect the amount of such background in the RS sample to be about twice the
number of events in the WS sample, since two tracks rather than one can be
misidentified as the candidate lepton.

To exclude feedthrough from the hadronic decays
$D^+ \rightarrow K^- K^+ \pi^+$
and  $D^+_s \rightarrow K^- K^+ \pi^+$, we exclude events in which the
invariant mass of the three charged particles (with the lepton candidate
interpreted as a pion) is consistent with the $D^+$ or $D^+_s$ mass. To
exclude feedthrough from  the more copious semileptonic decay
$D^+ \rightarrow \overline{K}^{\star 0}
\ell^+ \nu_\ell$, we exclude events in which the invariant mass of the 
two hadrons, with $K^-$ and $\pi^+$ masses assigned, is consistent
with $\overline{K}^{*\,0}(892)$.

Two kinematic signatures display the $D_s^+ \rightarrow \phi\, \ell^+\,
\nu_\ell$ signal: a $D_s$ peak in the minimum parent mass $M_{min}$ and
a $\phi$ peak in $K^+\,K^-$ invariant mass. $M_{min}$ is defined as the
invariant mass of $K K \ell \nu_\ell$ when the neutrino momentum component
along the $D^+_s$ direction of flight is ignored. The $M_{min}$ distributions
for the electron and muon channels are shown in the top left plots of Figs. 1
and 2, respectively; $D_s$ peaks are evident in the RS events.
We retain events with $M_{min}$ in the range 1.7-2.1 ${\rm GeV}/c^2$, as
indicated by arrows in the figures. The distribution of $K K$ invariant mass
for the retained events is shown in the top right plots of Figs. 1 and 2; the
RS sample exhibits a clear $\phi$ mass peak. Candidates with $1.014 < M_{K
K} < 1.026 \: {\rm~GeV}/c^2$ were retained, yielding final data samples of
166 right-sign and 11 wrong-sign events for the electron channel and 161
right-sign and 17 wrong-sign events
for the muon channel.

A Monte Carlo calculation simulated the response of the E791
detector to $D_s^+$ semileptonic decays.
A sample of events was
generated according to the differential decay rate (Eq.~22 in Ref.\cite{KS}),
with form factor ratios $r_2 = 0.78$, $r_V = 1.74$, and $r_3 =0$. The same
selection criteria were applied to the Monte Carlo events as to real data.
Out of 4.5 million events generated in the two modes, 13640 decays passed
all cuts. The bottom plots in Figs. 1 and 2 compare the background-subtracted
(RS$-2\times$WS) $\phi$ signals to the Monte Carlo $K^+ K^-$ spectra
from $D_s^+ \rightarrow \phi\, \ell^+\, \nu_\ell$ decays.
The agreement between the two distributions indicates that 
the background under the $\phi$ peak is adequately described by twice the WS
events.

The differential decay rate~\cite{KS} is expressed in terms of four
independent kinematic variables: the square of the momentum transfer ($q^2$),
the polar angle $\theta_V$  between the $K^+$ and $D^+_s$ in the $\phi$ rest
frame, the polar angle $\theta_\ell$  between the $\nu_\ell$ and $D^+_s$ in
the $W^+$ rest frame, and the azimuthal angle $\chi$ between the $\phi$ and
$W^+$ decay planes in the $D^+_s$ rest frame. 
Our definition of the polar angle $\theta_\ell$ is related to the
definition used in Ref.\cite{KS} by $\theta_\ell \to \pi - \theta_\ell$.
We do not use $\chi$ in this analysis because our data sample is not large
enough to profit from the small increase in sensitivity that $\chi$ can
provide.

Semileptonic decays cannot be fully reconstructed due to the undetected
neutrino. With the available information about the $D^+_s$ direction of flight
and the charged daughter-particle momenta, the neutrino momentum (and all the
decay's kinematic variables) can be determined up to a two-fold ambiguity if
the parent mass is constrained. Monte Carlo studies show that the differential
decay rate is more accurately determined, on average, if it is calculated
with the solution corresponding to the lower laboratory-frame neutrino
momentum.

To extract the form factor ratios, the distribution of data points in the
three-dimensional kinematic variable space is fit to the full expression for
the differential decay rate including lepton mass terms, but with $A_3$ set to
zero. We use an unbinned maximum-likelihood fitting technique~\cite{SMW}
in which the likelihood function is computed from the
density of Monte Carlo events near each data event
in the three-dimensional space of kinematic variables; the same Monte Carlo
events are reweighted to provide different trial values of form factor ratios.
To include background
in the fit, a similar likelihood function based on twice the density of
wrong-sign events around each right-sign event is used.

With the above method, the fitted results are subject to small systematic
biases which originate from two sources: (a) approximate normalization of the
likelihood function and (b) nonlinearity of the decay rate within the volume
centered on the data point. These systematic biases of the fitted parameters
were determined from Monte Carlo studies: $\delta r_V = +0.18 \pm 0.10$ and
$\delta r_2 = -0.12 \pm 0.05$ for the electron channel, and  $\delta r_V =
+0.14 \pm 0.09$ and $\delta r_2 = -0.10 \pm 0.06$ for the muon channel. After
correction for these biases by subtracting these $\delta r$\hspace{-0.01cm}'s
from the measured values, the final form factor ratios and their statistical
errors are $r_V = 2.24 \pm 0.47$ and  $r_2 = 1.64 \pm 0.34$ with a
correlation coefficient of $-0.15$  for the electron channel, and $r_V = 2.31
\pm 0.54$ and  $r_2 = 1.49 \pm 0.36$ with a correlation coefficient of $-0.20$
for the muon channel.

Fig.~3 compares the background-subtracted data in the kinematic variables
$q^2$, $\cos{\theta_\ell}$ and $\cos{\theta_V}$ for the electron and muon
channels with Monte Carlo predictions using our measured form factor ratios.
Note that although the $q^2$ distributions for the two channels are quite
different, due in large part to lepton-mass effects,  the form factor ratios
agree well.

The most important sources of systematic uncertainty and their estimated
contributions to the errors in $r_V$ and $r_2$ are listed in
Table~\ref{tabone}. The largest systematic effects are associated with small
densities of Monte Carlo events in kinematic variable space. We reject data
events (5 total) with zero nearby Monte Carlo events in this space; the
dependence of the form factor ratios on requiring more nearby Monte Carlo
events was studied by varying the minimum number from 1 to 5. Dependence of
$r_V$ and $r_2$ on the fineness of partitioning of the variable space was
investigated by varying this parameter by a factor of three. The systematic
biases in $r_V$ and $r_2$ described above were determined from ensembles of
Monte-Carlo experiments; the small uncertainties in them are largely
statistical. Sensitivity to possible inaccuracies in the Monte Carlo
simulation of detector response was estimated with several sets of reasonable
event-selection criteria.  Finally, sensitivity to uncertainty in the amount
of background was estimated by multiplying the wrong-sign events by weights
ranging from 0.5 to 3.5.

The consistency within errors of our results measured in the electron and
muon channels indicates that lepton mass effects are handled correctly in the
fits.
This agreement also
supports the assumption that strong interaction effects,
incorporated in the values of form factor ratios, do not depend on the
particular $W^+$ leptonic decay. We therefore combine the results measured for
the electron and muon decay modes in a weighted average:  $r_V =2.27 \pm 0.35
\pm 0.22$ and  $r_2 = 1.57 \pm 0.25 \pm 0.19$. The statistical and systematic
uncertainties of the averaged results were determined using the general
procedure described in Ref.\cite{LYONS} (Eqns.~3.40 and~3.40${}^{\prime}$).
The largest source of systematic error has a positive correlation between 
the two samples.

Table~\ref{tabtwo} and the top portion of Fig. 4 compare the values of the
form factor ratios $r_V$ and $r_2$ measured by E791 with previous
experimental results~\cite{CLEO,E653,E687}. The size of the data sample and
the decay channel are listed in Table~\ref{tabtwo} for each experiment. The
E791 combined results are consistent with the previous world
averages~\cite{PDG98} and have total errors 20\% - 30\% smaller than the
world averages.

Table~\ref{tabthree} and Fig.~4 (bottom) compare the combined E791 results
for $r_V$ and $r_2$ to published theoretical predictions for $D_s^+
\rightarrow \phi \, \ell^+ \, \nu_\ell$~\cite{BKS,LMMS,ISGW2} and to E791
measurements~\cite{renata,daniel} for $D^+ \rightarrow \overline{K}^{\,*0}\,
\ell^+ \,\nu_\ell$. Our value of $r_V$ for the $D_s$ decay is in good
agreement with all quoted predictions, and is also consistent with the E791
measurement for the $D^+$ decay. For $r_2$, our error for the $D_s$ decay is
significantly smaller than the spread in theoretical predictions.
Furthermore, our measurement for $D_s$ is more than two standard deviations
higher than the E791 $D^+$ result. This discrepancy in the measured value of
$r_2$ for $D_s$ and $D^+$ decays is even larger, about 3.3 standard
deviations, when new world averages are computed using data from this letter
and from Refs. \cite{daniel}~and \cite{PDG98}: 
$r_2=1.58\pm0.25$ for $D_s^+ \rightarrow \phi \, \ell^+ \, \nu_\ell$, and
$r_2=0.73\pm0.07$ for $D^+ \rightarrow \overline{K}^{\,*0}\,\ell^+ \,\nu_\ell$.
Since the quoted theoretical papers predict form factor ratios equal to
within 10\% for the two decays, it is difficult for these  predictions  to
agree  with both $D_s$ and $D^+$ data.

To summarize, we have measured the form factor ratios 
for the semileptonic decay $D^+_s \rightarrow \phi\, \ell^+ \nu_\ell$ 
in both electron and muon modes.
For the electron mode
 $r_V = 2.24 \pm 0.47 \pm 0.21$ and $r_2 = 1.64 \pm 0.34 \pm 0.20$, and 
for  the muon mode 
 $r_V = 2.31 \pm 0.54 \pm 0.26$ and $r_2 = 1.49 \pm 0.36 \pm 0.20$. 
The combined E791 results for  
 $D^+_s\rightarrow \phi\, \ell^+ \nu_\ell$ are
 $r_V = 2.27 \pm 0.35 \pm 0.22$ and $r_2 = 1.57 \pm 0.25 \pm 0.19$.
The $r_V$ result is consistent with the expected $SU(3)$ flavor symmetry
between $D_s$ and $D^+$ semileptonic decays; however, the $r_2$ result 
appears inconsistent with this expectation.

We gratefully acknowledge the assistance from Fermilab and
other participating institutions. This work was supported by the
Brazilian Conselho Nacional de Desenvolvimento Cient\'\i fico e
Technol\'{o}gico, CONACyT (Mexico), the Israeli Academy of Sciences and
Humanities, the U.S. Department of Energy, the U.S.-Israel Binational
Science Foundation, and the U.S. National Science Foundation.

\begin{small}
\begin{table}[t]
\caption{Main contributions to systematic uncertainties 
in the form-factor ratios for  $D^+_s \rightarrow \phi\, e^+ \nu_e$
and $D^+_s \rightarrow \phi\, \mu^+ \nu_\mu$.}
\label{tabone}
\begin{center}
\begin{tabular}{lcc}
{\bf Source of uncertainty}  
 & $\sigma_{r_V}\, (e,\mu)$ & $\,\,\sigma_{r_2}\, (e,\mu)$\\
 \hline
Min. \# of MC points per data point   & 0.16,$\,$0.19 & 0.17,$\,$0.13 \\
Size of partitions in variable space  & 0.06,$\,$0.11 & 0.06,$\,$0.10 \\
Systematic biases in fitting technique & 0.10,$\,$0.09 & 0.05,$\,$0.06 \\
Simulation of detector effects  & 0.06,$\,$0.08 & 0.06,$\,$0.07 \\ 
Number of background events & 0.04,$\,$0.05 & 0.03,$\,$0.04 \\
\hline
{\bf Total} 			 & 0.21,$\,$0.26 & 0.20,$\,$0.20 \\
\end{tabular}
\end{center}
\end{table}

\begin{table}[tbh]
\caption{Comparison of E791 results 
for  $D^+_s \rightarrow \phi\, \ell^+ \nu_\ell$
form factor ratios with previous experimental results.}
\label{tabtwo}
\begin{center}
\begin{tabular}{llll}
Exp. & Events & $r_V = V(0)/A_1(0)$ & $r_2 = A_2(0)/A_1(0)$  \\ 
\hline 
E791 &271 ($e+\mu)$ & $2.27 \pm 0.35 \pm 0.22$ & $1.57 \pm 0.25 \pm 0.19$ \\
E791 &144 ($e$) & $2.24 \pm 0.47 \pm 0.21$ & $1.64 \pm 0.34 \pm 0.20$ \\
E791 &127 ($\mu)$ & $2.31 \pm 0.54 \pm 0.26$ & $1.49 \pm 0.36 \pm 0.20$ \\
\hline
CLEO\cite{CLEO} & 308 ($e$) & $0.9 \pm 0.6 \pm 0.3$ & $1.4 \pm 0.5 \pm 0.3$ \\
E653\cite{E653} & 19 ($\mu$) & $2.3^{+1.1}_{-0.9} \pm 0.4$ & $2.1^{+0.6}_{-0.5} \pm 0.2$ \\
E687\cite{E687} & 90 ($\mu$) & $1.8 \pm 0.9 \pm 0.2$ & $1.1 \pm 0.8 \pm 0.1$\\
\hline
\multicolumn{2}{l}{Prev. world avg.~\cite{PDG98}} 
 & $1.5\pm 0.5$ & $1.6\pm 0.4$ \\
\end{tabular}
\end{center}
\end{table}
\vspace*{-0.1in}

\begin{table}[tbh]
\caption{
Comparison of E791  $D^+_s \rightarrow \phi\, \ell^+ \nu_\ell$ form factor
ratios with theoretical predictions 
for $D_s^+ \rightarrow \phi \, \ell^+ \, \nu_\ell$
and with E791 results for 
$D^+ \rightarrow \overline{K}^{\,*0}\, \ell^+ \,\nu_\ell$.}
\label{tabthree}
\begin{center}
\begin{tabular}{lll}
Group                &        $r_V$                  &       $r_2$       \\ 
\hline
E791 $D^+_s \rightarrow \phi\, \ell^+ \nu_\ell$     
  &    $2.27 \pm 0.41$      & $1.57 \pm 0.31$ \\ \hline
&&\\[-2.8mm]
BKS\cite{BKS}&$2.00\pm 0.19^{+0.20}_{-0.25}$&$0.78\pm 0.08^{+0.17}_{-0.13}$ \\    
LMMS\cite{LMMS}     &        $1.65 \pm 0.21$ &     $0.33 \pm 0.33$ \\
ISGW2\cite{ISGW2}&      $2.1$          &     $1.3$ \\
\hline
&&\\[-2.8mm]
E791 $D^+ \rightarrow \overline{K}^{\,*0}\, \ell^+ \,\nu_\ell$\cite{daniel}
&$1.87\pm 0.11$ & $0.73\pm 0.10$ \\ 
\end{tabular}
\end{center}
\end{table}
\end{small}

\begin{figure}[tbh]
\centerline{
\psfig{figure=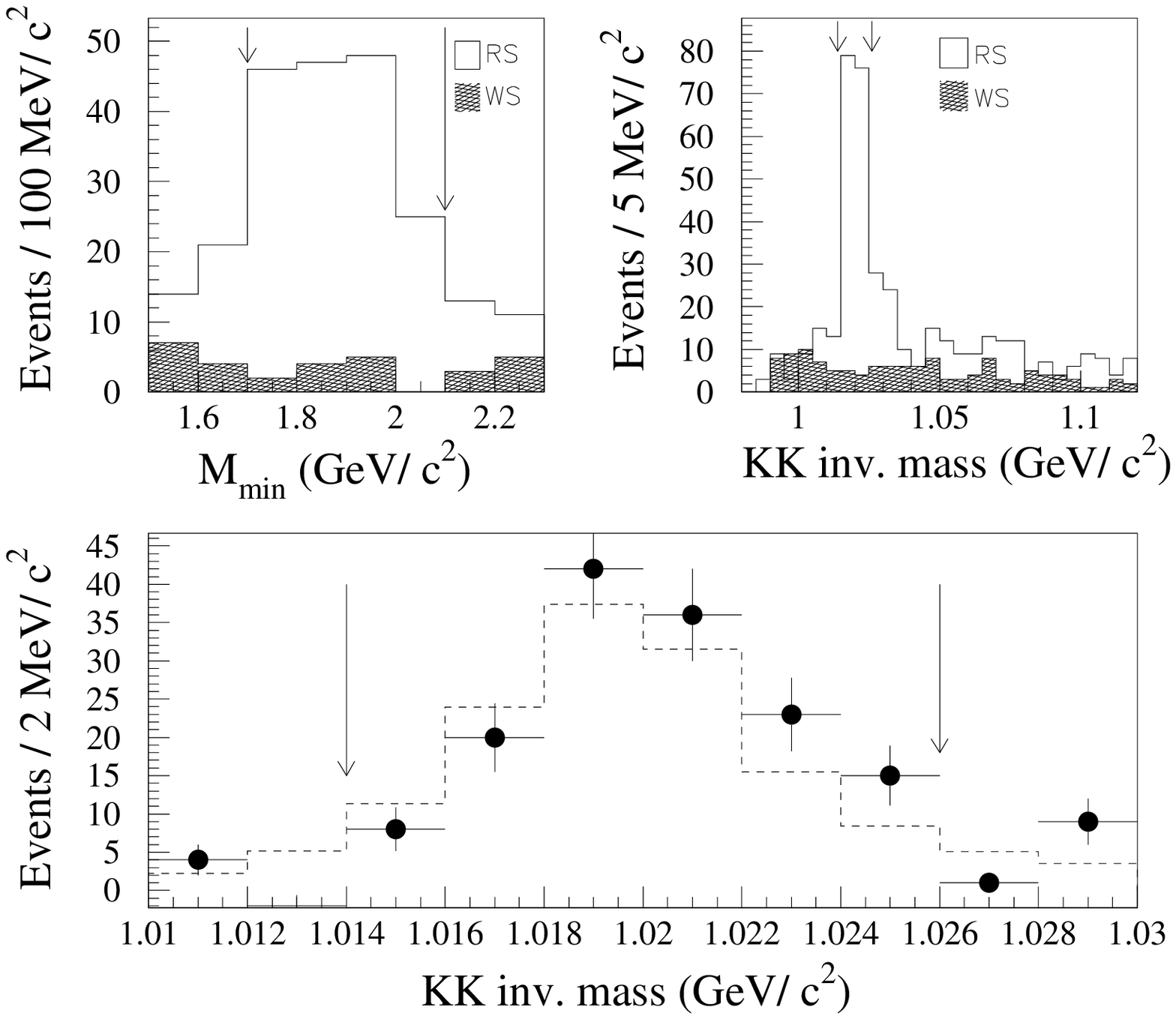,height=4.0in,width=4.0in}}
\caption{Distributions of minimum parent mass $M_{min}$ and $K K$ invariant
mass for $D^+_s \rightarrow \phi\, e^+ \nu_e$ candidate events.
Right-sign (RS) and wrong-sign (WS) samples are defined in the text.
Top left: $M_{min}$ for events with $K K$ mass in the range
1.014 to 1.026 GeV/$c^2$. Top right: $K K$ invariant mass
for events with $M_{min}$ in the range 1.7 to 2.1 GeV/$c^2$.
Bottom: background-subtracted 
(RS$-2\times$WS) $K K$ mass distribution (crosses)
compared to Monte Carlo prediction (dashed histogram) for
events with $M_{min}$ in the range 1.7 to 2.1 GeV/$c^2$ and $K K$ mass in the
range 1.014 to 1.026 GeV/$c^2$ .
All candidates pass all the other final selection cuts. The arrows indicate
the range of the final sample.}
\end{figure}

\begin{figure}[tbh]
\centerline{
\psfig{figure=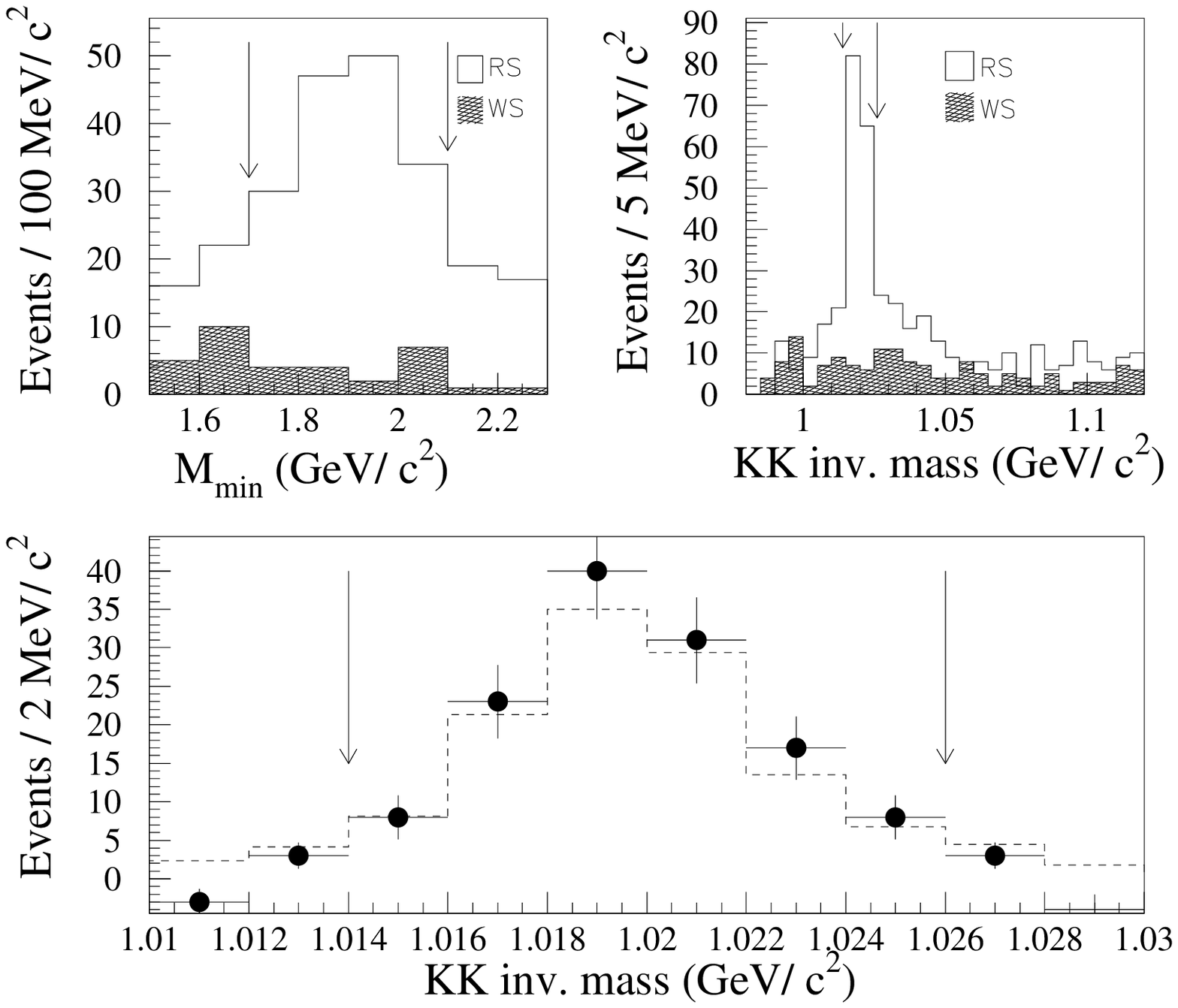,height=4.0in,width=4.0in}}
\caption{Distributions of minimum parent mass $M_{min}$ and $K K$ invariant
mass for $D^+_s \rightarrow \phi\, \mu^+ \nu_\mu$ candidate events.
Right-sign (RS) and wrong-sign (WS) samples are defined in the text.
Top left: $M_{min}$ for events with $K K$ mass in the range
1.014 to 1.026 GeV/$c^2$. Top right: $K K$ invariant mass
for events with $M_{min}$ in the range 1.7 to 2.1 GeV/$c^2$.
Bottom: background-subtracted 
(RS$-2\times$WS) $K K$ mass distribution (crosses)
compared to Monte Carlo prediction (dashed histogram) for
events with $M_{min}$ in the range 1.7 to 2.1 GeV/$c^2$ and $K K$ mass in the
range 1.014 to 1.026 GeV/$c^2$ .
All candidates pass all the other final selection cuts. The arrows indicate
the range of the final sample.}
\end{figure}

\begin{figure}[tbh]
\centerline{
\psfig{figure=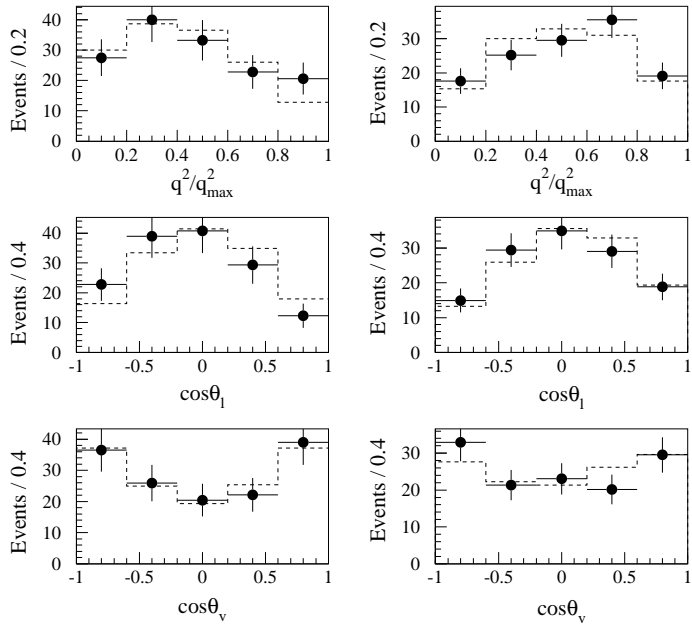,height=4.0in,width=4.0in}}
\caption{Comparison of kinematic variable distributions for
background-subtracted data (crosses) with Monte Carlo predictions (dashed
histograms) for electron (left) and muon (right) channels. The predictions
use our best-fit values for the form factor ratios.}
\end{figure}
\pagebreak

\begin{figure}
\centerline{
\psfig{figure=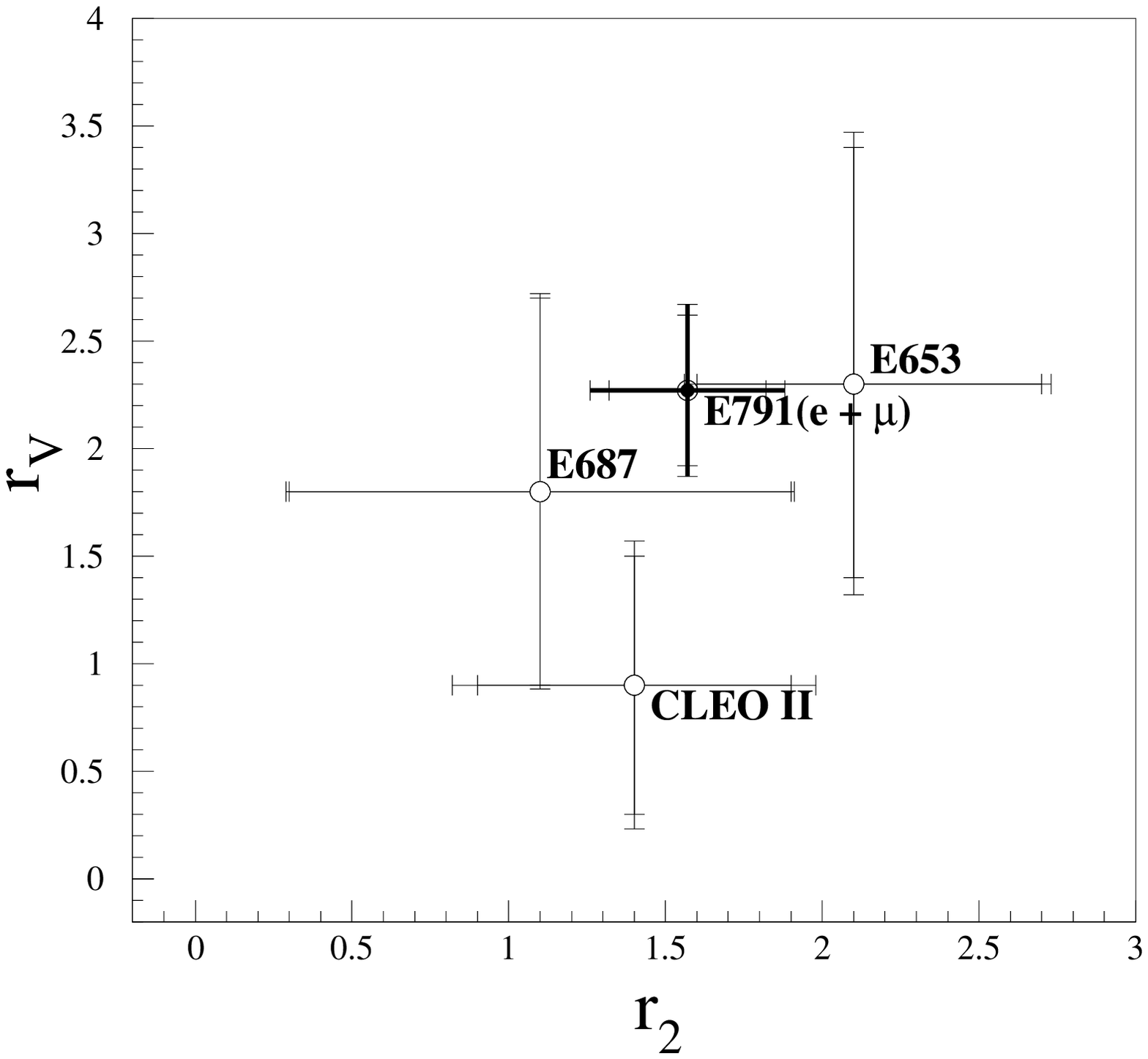,height=4.in,width=4.in}}
\vspace*{-0.5in}
\centerline{
\psfig{figure=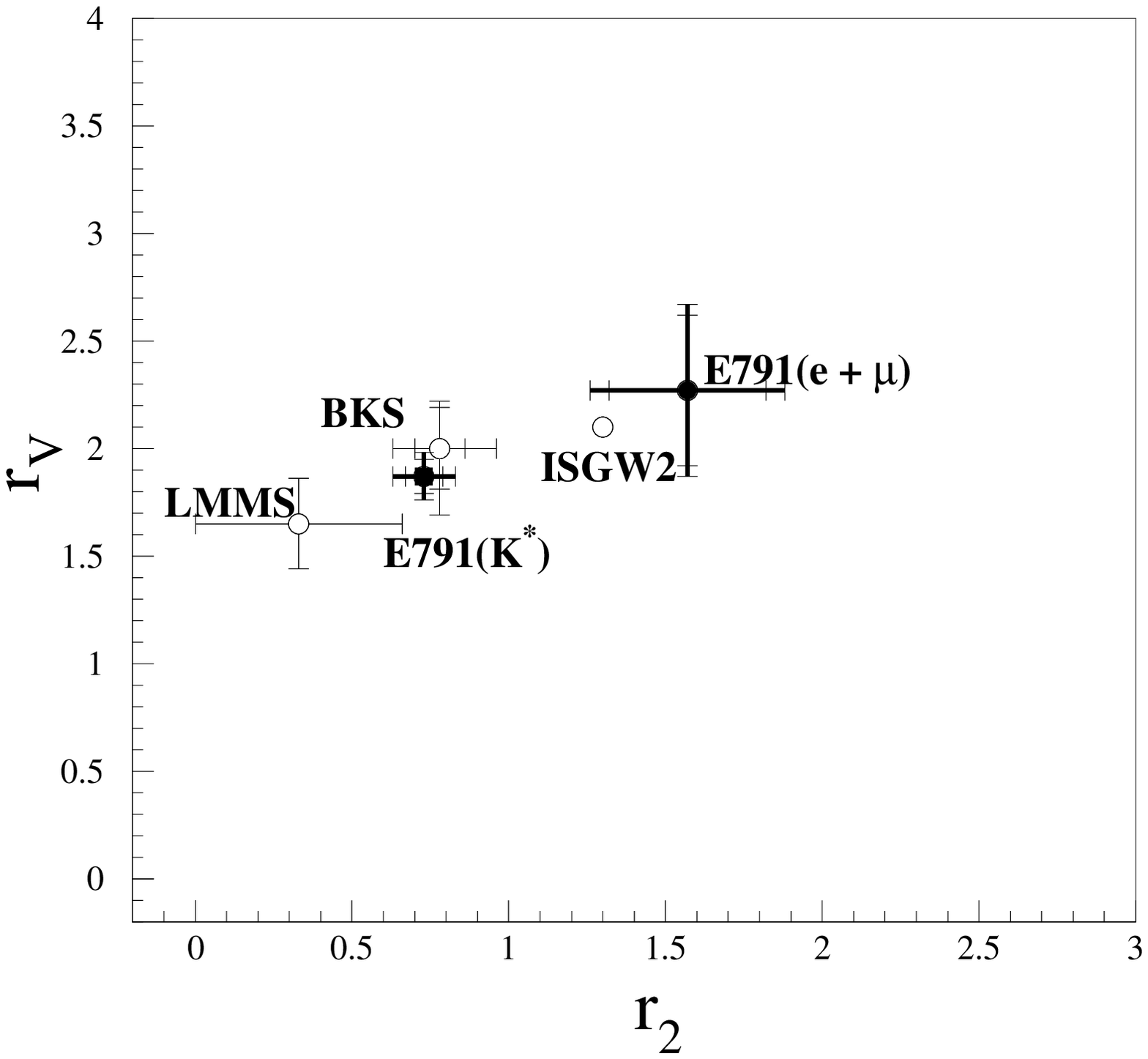,height=4.in,width=4.in}}
\caption{Top: Comparison of experimental measurements of form factor ratios
$r_V$ and $r_2$ for $D^+_s \rightarrow \phi\, \ell^+ \nu_\ell$. The smaller
error bars indicate the statistical errors and the larger ones show the
statistical and sytematic errors added in quadrature. Bottom: Comparison of
E791 measured form factor ratios for $D_s^+ \rightarrow \phi\, \ell^+\,
\nu_\ell$ with theoretical predictions 
for $D_s^+ \rightarrow \phi \, \ell^+ \, \nu_\ell$
and with E791 measurements for
$D^+ \rightarrow \overline{K}^{\,*0}\, \ell^+ \,\nu_\ell$.
}
\end{figure}


\begin{thebibliography}{99}
\vspace*{-0.3in}

\bibitem{KS} J.G. K\"{o}rner and G.A. Schuler, 
Phys. Lett. B~226 (1989) 185.

\bibitem{renata} Fermilab E791 Collaboration, E.M. Aitala et al., 
Phys. Rev. Lett. 80 (1998) 1393.

\bibitem{daniel} Fermilab E791 Collaboration, E.M. Aitala et al., 
Phys. Lett. B~440 (1998) 435.

\bibitem{KAMAL}
A.N. Kamal and A.B. Santra, Z. Phys. C~71 (1996) 101.


\bibitem{PDG94} R.J.~Morrison and J.D.~Richman, semileptonic decay
mini-review in
Particle Data Group, Review of Particle Physics, 
Phys. Rev. D~50 (1994) 1565.


\bibitem{APPEL} J.A. Appel, Ann. Rev. Nucl. Part. Sci. 42 (1992) 367;
  D.J. Summers et al., 
\uppercase\expandafter{\romannumeral27}  Rencontre de Moriond, Les
Arcs, France (15-22 March 1992) 417;
Fermilab E791 Collaboration, E.M. Aitala et al., 
xxx.lanl.gov e-print archive hep-ex/9809029 (1998), submitted to Phys. Rev. D.

\bibitem{SMW} D.M. Schmidt, R.J. Morrison, and M.S. Witherell, Nucl. Instrum.
Methods A~328 (1993) 547.

\bibitem{LYONS} L. Lyons, Statistics for Nuclear and Particle Physicists
(Cambridge University Press, Cambridge, 1986).

\bibitem{CLEO} 
CLEO Collaboration, P. Avery et al., Phys. Lett. B~337 (1994) 405.

\bibitem{E653} 
Fermilab E653 Collaboration, K. Kodama et al., Phys. Lett. B~274 (1992) 246.

\bibitem{E687} 
Fermilab E687 Collaboration, P.L. Frabetti et al., Phys. Lett. B~307 (1993) 262.

\bibitem{PDG98} 
Particle Data Group, Review of Particle Physics,
Eur. Phys. J. 3 (1998) 519.

\bibitem{BKS} C.W. Bernard, Z.X. El-Khadra, and A. Soni,  
Phys. Rev. D~45 (1992) 869.

\bibitem{LMMS} V. Lubicz, G. Martinelli, M. S. McCarthy, and C. T. Sachrajda,
Phys. Lett. B~274 (1992) 415.

\bibitem{ISGW2} D. Scora and N. Isgur,  Phys. Rev. D~52 (1995) 2783.
We have used the $q^2$ dependence assumed in our fits to extrapolate
the predicted form factors from $q^2_{max} = 0.90 {\rm{~(GeV}}/c)^2$
to $q^2=0$.

\end{thebibliography}
\end{document}